\documentclass[12pt,dvips]{article}

\usepackage{amsmath,amssymb,exscale}
\usepackage{array,multicol}
\usepackage{afterpage,float,flafter}
\usepackage{epsfig,rotating,pifont}
\usepackage{cite}
\input{epsf}
\@addtoreset{equation}{section} \makeatother

\begin{document}

\vskip 1cm

\begin{center}
{\Large \bf  Black Holes with Quantum Massive Spin-2 Hair }

\vskip 1cm {Gia Dvali\footnote{\it  email:  dvali@physics.nyu.edu}}

\vskip 1cm
{\it Center for Cosmology and Particle Physics, Department of Physics, New York University, New York, NY 10003}\\
\end{center}

\date{}

\begin{abstract}

 We show that black holes can posses a long range quantum mechanical hair associated with a  massive  spin-2 field, which can be detected by a  stringy  generalization of the Aharovon-Bohm effect, 
 in which a string loop lassoes the black hole. The long distance effect persist for arbitrarily high mass
 of the spin-2 field.  An analogous  effect is exhibited by a massive antisymmetric two-form field.  We make a close parallel between the two and  the ordinary Aharonov-Bohm phenomenon,  and also show that in the latter case the  effect can be experienced even by the electrically-neutral particles, provided some boundary terms are  added to the action.  
 
\end{abstract}

\newpage
\renewcommand{\thepage}{\arabic{page}}
\setcounter{page}{1}

\section{Introduction}

 Long time ago,  Bekenstein\cite{nohair} (see also Teitelboim\cite{nohair1}) proved that  black holes cannot maintain any  time-independent classical hair 
 of massive  fields, in particular,  a hair of a massive spin-2 meson.  In this paper we wish to show that although classical hair is absent, 
 nevertheless black holes can posses a quantum mecanical hair 
 under the massive spin-2 field. 
  This hair can be detected at infinity  by means of the stringy generalization of the 
 Aharonov-Bohm effect.  The similar statement is true about the massive antisymmetric Kalb-Ramond
 two-form (dual to a pseudoscalar axion).  
 Phenomenologically,  an interesting fact is that  the large distance effect 
 survives for an arbitrarily high  mass of  the hair-providing field.  
 
 Our work complements  the list of other  possible quantum-mechanical hairs suggested earlier.  
   Possibility of a quantum mechanical hair under the discrete gauge symmetries was pointed out
   in\cite{discrete}. Such hair can be detected by the usual  Aharonov-Bohm effect, provided  there are 
   cosmic strings carrying the flux of the magnetic field of the broken $U(1)$\cite{ABmassive}. 
   
    The authors of \cite{pseudo} discovered black holes with the hair under the massless axion field, and showed that it can be detected by the stringy Aharonov-Bohm effect.  Extension of the solution for  the massive axion was 
   demonstrated in\cite{massivepseudo}.  
   In the second half of the paper we show that, although the long distance effect persists for an arbitrarily heavy two-form, when applied to the QCD axion, the issue of detectability becomes subtle.
  
   We shall establish a close parallel between the spin-2 and the ordinary  abelian Aharonov-Bohm effects, and as a byproduct show that the ordinary Aharovon-Bohm effect in the presence of the photon mass can be experienced also by the electrically neutral particles, provided the certain boundary terms are included in the action.  
   
   Because our arguments are topological, they must apply to arbitrarily small black holes 
 for which the usual quasi-classical treatment breaks down.  In particular, they could be useful in 
 tracing the latest stages of the black hole evaporation, and in understanding the interactions between the  microscopic black holes and QCD or the fundamental strings.  
    
 \section{Spin-2 Quantum  Hair}
  
 Following  Bekenstein, let us consider a linearized theory of a massive spin-2 particle.   On a (asymptotically) flat background, there is an unique ghost-free theory of this sort, Pauli-Fierz 
theory,  described by the following equation  
\begin{equation}
\label{pf}
 \mathcal{E}^{\alpha\beta}_{\mu\nu} h_{\alpha\beta}\, 
\, - \, m^2\, (h_{\mu\nu} \, - \, \eta_{\mu\nu} h) \,  = \, T_{\mu\nu}. 
\end{equation}
 Where,
\begin{equation}
\label{einstein}
\mathcal{E}^{\alpha\beta}_{\mu\nu} h_{\alpha\beta}\, = \, \square h_{\mu\nu} \, - 
\, \square \eta_{\mu\nu} h  \, - \, \partial^{\alpha}\partial_{\mu} h_{\alpha\nu} \, -\, 
\partial^{\alpha} \partial_{\nu} h_{\alpha\mu} \, + \, 
\eta_{\mu\nu} \partial^{\alpha}\partial^{\beta}h_{\alpha\beta}\, + \, \partial_{\mu}\partial_{\nu} h
\end{equation}
is the linearized Einstein's tensor, and as usual $h \equiv  \eta_{\alpha\beta}h^{\alpha\beta}$.
$T_{\mu\nu}$ is a conserved source, which we shall set to zero everywhere, except 
possibly at the black hole singularity.  
The Newton's coupling is set to one, for simplicity.  

 
The generalization to an arbitrary background is, as usual,  through substituting all the derivatives 
by the covariant ones and replacing $\eta_{\mu\nu}$ by the background metric $g_{\mu\nu}$. 

 Note that the  linearized Einstein term (\ref{einstein}) is invariant  under the following gauge transformation
\begin{equation}
\label{gaugeh}
h_{\mu\nu} \, \rightarrow h_{\mu\nu}  \, +  \, \partial_{\mu}\xi_{\nu} \, + \, \partial_{\nu} \xi_{\mu},
\end{equation}
This invariance is, however, (seemingly) not supported by the mass term in (\ref{pf}).
  As a result of this, $h_{\mu\nu}$ becomes an observable quantity.   
 Bekenstein then showed that  for the Schwarzschild  background metric,  $h_{\mu\nu}$ has to vanish 
 identically everywhere outside the horizon.   The essence of the proof relies on the fact that in the 
 massive theory  $h_{\mu\nu}^2$ is a physical scalar, and thus must be regular at the horizon,
 which only can be accommodated if  $h_{\mu\nu}$ is zero. Since we are not questioning this proof, we shall not repeat it here.  

 Instead,  we wish to show that, although the classical  hair vanishes,  black hole nevertheless can support a 
 quantum mechanical hair, that can be detected at infinity.  The possibility of the quantum hair 
 relies on the {\it hidden} gauge invariance exhibited by the Pauli-Fierz equation (\ref{pf}). 
 Although with the addition of the mass term $h_{\mu\nu}$ becomes observable, 
 the gauge invariance is nevertheless not lost.  The point is that the gauge invariance is a redundancy 
 of the description, and persist even in the massive phase. 
 
  To understand this, we should remember that addition of the mass term has changed the theory 
  discontinuously, by increasing the number of the physical degrees of freedom residing in $h_{\mu\nu}$ from   $2$ to $5$\cite{vDVZ}.   So talking about the gauge invariance is meaningless, unless we specify how the 
  different degrees of freedom transform under it. In the present case, the answer is that  the addition of the mass
  term does not jeopardize the gauge invariance, since the $3$ new degrees of freedom 
  exactly compensate the transformations of the other  two, rendering the full $h_{\mu\nu}$,  and thus the mass term,  gauge invariant.

 To make this clear,  
we shall first rewrite 
equation (\ref{pf}) in the manifestly gauge invariant form, introducing extra degrees of freedom 
in form of the St\"uckelberg fields. 
To do this,  we recast  $h_{\mu\nu}$ in the following way 
\begin{equation}
\label{vector}
h_{\mu\nu} \, = \, \hat{h}_{\mu\nu} \, + \, \partial_{\mu}A_{\nu} \, + \, \partial_{\nu} A_{\mu},
\end{equation}
where the St\"uckelberg field $A_{\mu}$ is the massive vector that carries extra polarizations. 
Written in terms of $\hat{h}_{\mu\nu}$ and $A_{\mu}$, 
  \begin{equation}
\label{pfs}
 \mathcal{E}^{\alpha\beta}_{\mu\nu} \hat{h}_{\alpha\beta}\, 
\, - \, m^2\,  (\hat{h}_{\mu\nu} \, - \, \eta_{\mu\nu} \hat{h}  \, + \partial_{\mu}A_{\nu} \, + \, \partial_{\nu} A_{\mu} \, - \, 2 \eta_{\mu\nu} \partial^{\alpha}A_{\alpha}) \,  = \, 0
\end{equation}
the Pauli-Fierz equation (\ref{pf}) becomes manifestly invariant under the following gauge transformations 
\begin{equation}
\label{gauge}
\hat{h}_{\mu\nu} \, \rightarrow \hat{h}_{\mu\nu}  \, +  \, \partial_{\mu}\xi_{\nu} \, + \, \partial_{\nu} \xi_{\mu},
~~~~~~~A_{\mu} \, \rightarrow \, A_{\mu} \,  - \, \xi_{\mu}.
\end{equation}
Note that the Einstein's tensor is unchanged under the replacement (\ref{vector}),  due to its gauge invariance. 
The equation of motion for  $A_{\mu}$ has the following form. 
  \begin{equation}
\label{Aequ}
\partial^{\mu} F_{\mu\nu} \, = \, - \, \partial^{\mu}  (\hat{h}_{\mu\nu} \, - \, \eta_{\mu\nu} \hat{h} ),
\end{equation}
where $F_{\mu\nu} \, = \, \partial_{\mu}A_{\nu}  - \partial_{\nu}A_{\mu}$. 

 We shall now show that there is a topologically nontrivial spherically symmetric configuration, for
 which $h_{\mu\nu}$ is zero everywhere (away from the black hole singularity).  This is the configuration for which  $F_{\mu\nu}$ has a form of the magnetic field of a Dirac monopole
 placed at the center, with
 $\hat{h}_{\mu\nu}$ satisfying 
 \begin{equation}
\label{hmonopole}
 \hat{h}_{\mu\nu} \, = - (\, \partial_{\mu}A_{\nu} \, + \, \partial_{\nu} A_{\mu}). 
\end{equation}
The latter equality guarantees that $h_{\mu\nu}$ is  identically zero. 
Thus, In order to  produce a topologically non-trivial structure  we are mapping the magnetic monopole, with the topology $R_2\times S_2$ on the Schwarzschild metric with the same topology. 

 As it is well known,  in the presence 
of  the magnetic monopole $F_{\mu\nu}$ can be written as an exterior derivative of $A_{\mu}$
locally everywhere, but not globally.  This fact is enough  to guarantee the vanishing of  $h_{\mu\nu}$. 
 For instance, we can choose in the spherical coordinates,  
 \begin{equation}
\label{amonopole}
A_{\phi} \, = \, \mu \, {1 \, - \cos(\theta) \over r \sin(\theta)},~~~A_{\theta} \, =\, A_r\, = \, 0, 
\end{equation}
corresponding to a spherically symmetric radial magnetic field
\begin{equation}
\label{magnet}
\vec{\mathcal{M}} \, = \, \mu \, {\vec{r} \over r^3},
\end{equation} 
where $\mu$ is the magnetic charge of the monopole.  In this gauge, the Dirac string coincides 
with the negative $z$ semi-axis. Unobservability of the  Dirac string is guaranteed  
by the fact that there are no electrically charged particles under $A_{\mu}$, as demanded by the 
gauge symmetry. Indeed, any interaction of the sort
\begin{equation}
\label{aparticle}
 \int dX^{\mu}A_{\mu} ,
\end{equation} 
where $X^{\mu}$ are the particle coordinates,  is forbidden by gauge symmetry (\ref{gauge}). 

 As in the case of the Dirac magnetic monopole, we can describe the above configuration without any reference to the string singularity.  We can define the smooth  vector potentials  
on the two hemispheres in the following way  \cite{hemispheres}
\begin{eqnarray}
A_{\phi}^U & = & \mu {1 \, - \, cos \theta \over r  sin\theta}~~~~~~~~~0 \, < \, \theta \, < {\pi \over 2}  \\
A_{\phi}^L & = & - \, \mu {1 \, + \, cos \theta \over r  sin\theta}~~~~~~ {\pi\over 2}  \, < \, \theta \, < \pi 
\end{eqnarray}
In the same time we define $h_{\mu\nu}^U, h_{\mu\nu}^L$ according to (\ref{hmonopole}). 
Because $A^U$ and $A^L$ at the equator differ  by a single-valued gauge transformation, 
so do $h^U$ and $h^L$
\begin{equation}
\label{hul}
(h_{\mu\nu}^U \, - \, h_{\mu\nu}^L)|_{\theta= {\pi\over 2}} \, = \, 2\mu \partial_{\mu}\partial_{\nu}\phi\, |_{\theta = {\pi\over 2}},  
\end{equation}
and thus, they  describe the same physics.  

 If $A_{\mu}$ where an ordinary $U(1)$ gauge field, the Dirac quantization condition would follow from the requirement that the phase of an elementary particle be single valued around the equator. 
 As said above, such particles are absent in our case, by gauge invariance.  Which is the same argument that implied undetectability  of the Dirac string. 

In fact, not only the Dirac string, but
the whole configuration  is undetectable classically, because locally everywhere 
$\hat{h}_{\mu\nu}$ has a {\it pure gauge}  form.  However, due to the non-trivial topology, it can be
detected quantum mechanically  via the stringy generalization of the Aharonov-Bohm experiment.
Such a generalization of the Aharonov-Bohm effect was first discussed in\cite{pseudo} for the massless axion field.  
In order to design  such a measurement in our case, one needs the existence of strings to which $F_{\mu\nu}$ could couple. 
Strings may either be fundamental or of the solitonic type, such as cosmic strings. The precise microscopic 
nature is unimportant at large distances.  
 The effective gauge invariant coupling to the string has the following form 
\begin{equation}
\label{fstring}
q\, F_{\mu\nu}  \int d^2\sigma \partial_aX^{\mu}\partial_bX^{\nu} \epsilon^{ab}\delta^4(x - X),
\end{equation}
where $\sigma^a$ are the string world sheet coordinates, and $X^{\mu}(\sigma)$ are the embedding 
coordinates.  In the absence of a topologically non-trivial configurations, the coupling 
(\ref{fstring}) simply reduces to a  boundary term.  In the presence of a monopole, however, 
it leads to a nontrivial scattering, which can be observed in the string interference experiment. 
Indeed for any string that lassoes  the black hole, the action changes by 
\begin{equation}
\label{changes}
\Delta S \, = \, 4\pi q\mu,
\end{equation} 
which can be detected by interfering such a string with the one that avoids the black hole.


The effect will lead to an observable scattering unless $\mu$ and $q$  satisfy the quatization condition,
$ q\mu \, = \, {n \over 2} $,  which {\it a priory}  there is no reason for, since the black hole could carry an arbitrary quantum Spin-2 charge. 

 To summarize,  in the presence of strings, the Pauli-Fierz action can be represented in the following
form
\begin{equation} 
\label{total}
\hat{h}^{\mu\nu} \mathcal{E}^{\alpha\beta}_{\mu\nu} \hat{h}_{\alpha\beta}\, 
\, - \, m^2\,  \left ((\hat{h}_{\mu\nu} \, +  \partial_{\{\mu}A_{\nu\}})^2 \, - \, (\hat{h} \, + \, 2 \partial^{\alpha}A_{\alpha})^2 \,  \right)\, + \, 
q\, F_{\mu\nu}  \int d^2\sigma \partial_aX^{\mu}\partial_bX^{\nu} \epsilon^{ab}\delta^4(x - X),
\end{equation}
 which continues to describe a massive spin-2 particle with five degrees of freedom.  Black holes
 in this theory are labeled by an additional quantum mechanical spin-2 charge $\mu$, which is 
 undetectable classically, but is observable quantum mechanically. 

 An alternative derivation of  the quantum spin-2 hair, is  to start from (\ref{total}) and consistently integrate  out the $A_{\mu}$ field. The resulting effective theory only contains the two 
 helicity-2  states and is formulated exclusively in terms of the gauge-variant part of $h_{\mu\nu}$, without any compensating fields.  The locally-pure-gauge nature of the hair in this language becomes
 in a certain sense  more transparent.  For completeness, we  give such a treatment in the appendix
 both for massive spin-2 and massive spin-1 fields.

 Phenomenologically, the most interesting thing about the above quantum effect is that it persists for an 
arbitrarily large graviton  mass.  This opens up a possibility to measure effects mediated by   the heavy quantum fields at arbitrarily large distances.  Such a non-decoupling is characteristic
 to the ordinary Aharonov-Bohm effect, which is  known to also persist  in case of non-zero 
 photon mass \cite{ABmassive}. 
 
 \section{Parallel with the Aharonov-Bohm Effect in Massive $U(1)$}
 
  In order to make an useful parallel between the photon and the spin-2 cases, consider an usual Aharonov-Bohm effect\cite{AB} in the presence of the photon mass. The setup consists of an abelian massive gauge field 
$Z_{\mu}$  (an analog of  massive spin-2, $h_{\mu\nu}$) in the background of a magnetic flux 
localized at $\rho \, = \, 0$ (in the polar coordinates $\rho, \phi, z$). The magnetic flux plays the  role 
analogous to the black hole singularity, to which the long distance probes cannot  penetrate.  At low energies the dynamics of the massive photon is described by the Proca  Lagrangian 
\begin{equation}
\label{proca}
L\, = \, -{1 \over 4} F_{\mu\nu}F^{\mu\nu} \, + \, {1 \over 2} m^2\, Z_{\mu}Z^{\mu}.
\end{equation} 
If the microscopic origin of the mass is the Higgs effect, then $m^2$ is the expectation value of a dynamical heavy field.  In the latter case the flux at $\rho \, = \, 0$ can have a topological origin, and be 
represented by an  Abrikosov-Nielsen-Olesen vortex.  In the latter case Higgs will vanish at  $\rho=0$ due to the topological reasons. 
However, we are only interested in the interaction with the probe particles at  large distances,  where $m^2$ can be safely treated as constant. 

Now, from the simple energy considerations it is obvious that $Z_{\mu}$ must fall-off  rapidly  for 
$\rho \gg m^{-1}$, and one may think naively that the effect should disappear, because unlike the massless case, not only magnetic field, but also $Z_{\mu}$ itself becomes zero outside the flux region. 
This however is not the case, since the introduction  of the  mass term automatically adds an additional 
degree of freedom.  As in the massive spin-2 case, we can make this new degree of freedom explicit  by writing 
\begin{equation}
\label{zrep}
Z_{\mu} \, = \, \hat{Z}_{\mu} \, + \, \partial_{\mu} a,
\end{equation} 
where $a$ is the St\"uckelberg field, which in the Higgs case is an eaten-up Goldstone boson. 
It is easy to notice that $\hat{Z}_{\mu}$ and $a$ play the roles analogous to $\hat{h}_{\mu\nu}$ and 
$A_{\mu}$ respectively. Using the above representation, we can rewrite the Proca Langangian in 
the following form, 
\begin{equation}
\label{procag}
L\, = \, -{1 \over 4} F_{\mu\nu}F^{\mu\nu} \, + \, {1\over 2} \,m^2(\hat{Z}_{\mu} \, + \, \partial_{\mu}a)^2, 
\end{equation} 
which is manifestly gauge invariant under the transformations
\begin{equation}
\label{zgauge}
\hat{Z}_{\mu} \, \rightarrow \hat{Z}_{\mu}  \, +  \, \partial_{\mu}\omega
~~~~~~a \, \rightarrow \, a \,  - \, \omega,
\end{equation}
where $\omega$ is a transformation parameter. 
 Although, $Z_{\mu}$ vanishes outside the flux, individually $\hat{Z}_{\mu}$ and $\partial_{\mu}a$
 do not, and we have 
\begin{equation}
\label{zhata}
\hat{Z}_{\phi} \, =  \, - \,{\partial_{\phi} a \over \rho}  \, = \,  {1 \over \rho}.
\end{equation}
For simplicity we have taken an unit magnetic flux. 
 Now it is obvious that Aharonov-Bohm effect will persist, provided there exist some fractionally charged 
 particles coupled to $\hat{Z}_{\mu}$ and {\it not} to $Z_{\mu}$.  The latter situation  
 is usually taken for granted when the mass is generated via the Higgs mechanism.  However, 
 it is important to kate into the account that this is not the only option.  
    In fact,  the strict condition for having an observable effect is that the 
 particles couple to an arbitrary  combination $\alpha Z_{\mu} \, + \, \beta \partial_{\mu}a$, such that 
 $(\alpha  -  \beta)$ is  non-integer.  The closest analogy to (\ref{fstring}) would be to set 
 $\alpha\, = \, 0$ and only consider the following coupling  
\begin{equation}
\label{acoupling}
\beta \int \, dX^{\mu} \partial_{\mu}a,  
\end{equation}
where $X^{\mu}$ are  the particle coordinates. 
In a full analogy with (\ref{fstring}),  the coupling (\ref{acoupling}) also amounts to a boundary term, irrelevant for topologically-trivial configurations.  However, for a non-zero winding of the phase 
$\oint \partial a\, = \, 2\pi$,  when a particle is transported around the flux, the action changes by 
\begin{equation}
\label{sdelta1}
\Delta S \, = \, 2\pi \beta,
\end{equation}  
leading to the detectable effect for non-integer $\beta$.  The {\it per se} - interesting fact about the latter 
case, is that the particle exhibiting the Aharonov-Bohm effect  carries a {\it zero}  charge under the gauge group, since $\alpha \, = \, 0$.  This goes in contrast to usual wisdom about  the  Aharonov-Bohm  effect. 
 
 Thus, from the above discussion  the following lesson emerges.  In the presence of the photon mass, 
the susceptibility of a particle to the Aharonov-Bohm interaction  is not determined  by the 
Coulomb type  electric charge of a particle\footnote{Of course, for the massive photon there is no Coulomb 
electric flux at infinity. So under the Coulomb charge we mean the charge that determines the 
strength of the electric field 
at short distances. That is, the coupling constant  that would give  the usual Coulomb charge
in the sense of the Gaussian integral at infinity  for $m \rightarrow 0$.}.
Indeed, because the term (\ref{acoupling}) does not affect the equation of motion, the electric field of a particle coupled to the combination $\alpha Z_{\mu} \, + \, \beta \partial_{\mu}a$ is 
\begin{equation}
\label{electric}
\vec{E} \, = \, \alpha {\vec{r} \over r^3} \, e^{-mr}\, ( 1 \, + \, rm) 
\end{equation}  
and is determined by $\alpha$ only.  In the same time the Aharonov-Bohm effect is determined by the 
combination $\alpha - \beta$. If the latter is non-integer, the effect is detectable, even if $\alpha$ is an integer, and vise versa.  

 The above statement can be made in a more conventional language of the Higgs completion of the Proca theory.  
In this case the photon mass is generated by an expectation value of a complex scalar field 
$\Phi\, = \, |\Phi| e^{i a}$, where $\langle |\Phi|\rangle \, = \, m$.  The photon mass term is then replaced by   
\begin{equation}
\label{massreplace}
 m^2Z_{\mu}Z^{\mu} \, ~~\rightarrow \,~~ \mathcal{D}_{\mu}\Phi^*\mathcal{D}^{\mu}\Phi,
\end{equation} 
where,  as usual $\mathcal{D}_{\mu} \, \equiv \, \partial_{\mu} \, + \, i\, \hat{Z}_{\mu}$.   Thus,  the correct Proca limit of the Higgs theory is 
  \begin{equation}
\label{ptoh}
\Phi^* \mathcal{D}_{\mu}\Phi \, ~~\rightarrow \,~~ m^2Z_{\mu}.
\end{equation}

Now it is clear that in the presence of the photon mass, one can distinguish three types of possible 
interactions with the fermions (call them $\psi$).    The first category  are the fermions that are 
electrically-charged in the usual sense, and are coupled to the photon through  the minimal gauge invariant coupling,
\begin{equation}
\label{first}
\bar\psi \mathbf{D}_{\mu} \gamma^{\mu} \psi \,~~ \rightarrow \,~~ \int dX^{\mu} \hat{Z}_{\mu}, 
\end{equation}
where, $\mathbf{D} \, \equiv \, \partial_{\mu} \, + \, i\,\alpha \hat{Z}_{\mu}$. Such a fermion can only experience the Aharonov-Bohm effect if $\alpha$ is fractional.

The fermions of the second category are the ones that couple to the photon through the following gauge invariant coupling 
\begin{equation}
\label{second}
{-i \over 2M^2} \, (\Phi^*\mathcal{D}_{\mu} \Phi \, - \, h.c.) \, \bar\psi \gamma^{\mu} \psi \,~~ \rightarrow \,~~ 
{m^2 \over M^2} \, \int dX^{\mu} (\hat{Z}_{\mu} \, + \, \partial_{\mu}a), 
\end{equation}
where $M$ is a cutoff.  Such fermions will never experience the  Aharonov-Bohm effect,  although they do produce an electric field measured by the charge $m^2/M^2$, which can be fractional.  
The reason is that  the winding of $\hat{Z}_{\mu} \, + \, \partial_{\mu}a \equiv Z_{\mu}$ is zero.  

Finally, there are fermions that couple only to the longitudinal photons
\begin{equation}
\label{third}
\beta\, (\partial_{\mu} a) \, \bar\psi \gamma^{\mu} \psi \,~~ \rightarrow \,~~  \beta\, \int dX^{\mu} \partial_{\mu}a. 
\end{equation}
Although, such fermions do not produce any electric field, they nevertheless 
experience the effect for non-integer $\beta$.  The coupling (\ref{third}) can be written in terms of the Higgs field as 
\begin{equation}
\label{higgsterm}
{-i \beta \over \ 2 |\Phi|^2} \, (\Phi^*\partial_{\mu} \Phi \, + \, h.c.)  \, \bar\psi \gamma^{\mu} \psi,
\end{equation}
The coupling (\ref{third}) is singular at the origin, and singularity must be resolved by some microscopic 
physics.  However, this physics cannot affect our conclusions, since the effect is experienced  by the probe particles that never penetrate to the origin.  Also, in the case of spin-2 hair, the singularity 
is hidden behind the balck hole horizon.

 \section{Detectability of the Massive Axion Hair} 
 
 A very similar  Aharonov-Bohm type non-decoupling is also true for the quantum mechanical hair of a  massive antisymmetric two-form field $B_{\mu\nu}$. The massless case was considered in \cite{pseudo}, and persistence of the solution in the massive case was pointed 
 out in \cite{massivepseudo}.   We shall consider the Aharovon-Bohm  detectability
 of the massive case, and its relevance for the QCD axion.  
 
   Let us first review the massless case.  
 The lagrangian of a massless  antisymmetric Kalb-Ramond two-form 
 field $B_{\mu\nu}$ is 
  \begin{equation}
\label{RB}
H_{\alpha\beta\gamma} \,H^{\alpha\beta\gamma},
\end{equation}
where 
 \begin{equation}
\label{FB}
H_{\alpha\beta\gamma}\, = \, \partial_{[\alpha} B_{\beta\gamma]} \, = \, dB
\end{equation}
is the three-form field strength, and $d$ is the exterior derivative. 
This action is  invariant under  the following gauge symmetry  
\begin{equation}
\label{shiftB}
B_{\alpha\beta} \, \rightarrow  \, B_{\alpha\beta}  \, + 
\partial_{[\alpha}\xi_{\beta]},
\end{equation}
where $\xi_{\beta}$ is an arbitrary vector  (one-form). The authors of \cite{pseudo} then showed 
that there is a black hole solutions that carries a non-trivial $B_{\mu\nu}$-charge,
\begin{equation}
\label{bmonopole}
B_{\mu\nu} \, = \, \mu {\epsilon_{\mu\nu} \over r^2},  
\end{equation}
where $\epsilon_{\mu\nu}$ is an induced volume element of a two-sphere enclosing the
Schwarszchild solution.  Such a field is locally representable in form of an exterior derivative $B = d\xi$. Were we can choose
the gauge parameter $\xi_{\mu}$ to have the following asymptotic  form  
\begin{equation}
\label{ximonopole}
\xi_{\phi} \, = \, \mu \, {1 \, - \,  cos\theta \over r sin\theta}, ~~~\xi_{\theta} \, = \, \xi_r\, = 0.
\end{equation}
Because, $B$ has a locally-pure-gauge form, the field strength $H$ is identically zero everywhere, 
except the black hole singularity. The field however can be detected by an Aharonov-Bohm type effect,
provided $B_{\mu\nu}$ couples to strings, 
\begin{equation}
\label{bstring}
q\, B_{\mu\nu}  \int d^2\sigma \partial_aX^{\mu}\partial_bX^{\nu} \epsilon^{ab}\delta^4(x - X).
\end{equation}
The consideration about the non-zero scattering amplitude is identical to the one we gave for the 
gravity case.  

It is easy  to show that the solution  persists for the massive $B_{\mu\nu}$ field\cite{massivepseudo}. 
Our experience with the massive spin-2 field tells us how to proceed.  We first add a mass term 
to the Lagrangian  (\ref{RB}),
  \begin{equation}
\label{RBmass}
H_{\alpha\beta\gamma} \,H^{\alpha\beta\gamma} \, + \, m^2 B'_{\mu\nu}B^{'\mu\nu},
\end{equation}
where we have put a prime in order to distinguish the massive $B'_{\mu\nu}$ field from the massless one, $B_{\mu\nu}$.  Seemingly, just as in the case of spin-2, the mass term breaks the gauge invariance
(\ref{shiftB}).  And just as in that case, this is a false impression, because the  addition of a mass term
adds new degrees of freedom, which compensate gauge shift of 
$B_{\mu\nu}$ and make the mass-term invariant.  To make this explicit, we rewrite 
theory in a manifestly gauge invariant way, by  setting $B_{\mu\nu}' \, = \, B_{\mu\nu} \, + \, 
\partial_{[\alpha} A_{\beta]}$,  where $A_{\mu}$ is the St\"uckelberg field that 
under (\ref{shiftB}) shifts as (\ref{gauge}).  Now it is obvious,  that  (\ref{bmonopole}) persist to be a solution for $m \neq 0$, provided $A_{\mu}$ takes the form (\ref{amonopole}).   Since the 
stringy Aharonov-Bohm effect  only cares about the topologically non-trivial configuration of 
$B_{\mu\nu}$, it continues to be a detectable effect for an arbitrarily large  $m$. 

 However, just as in the case of the massive photon, detectability requires certain condition 
 to be satisfied,  not presented in the massless case.  In order for the effect to be observable, the strings
 should couple to a combination 
\begin{equation}
\label{bstring}
(\alpha B_{\mu\nu}  \, + \, \beta F_{\mu\nu})  \int d^2\sigma \partial_aX^{\mu}\partial_bX^{\nu} \epsilon^{ab}\delta^4(x - X)
\end{equation}
 such that, $ \mu(\alpha - \beta) \neq {n \over 2} $.

  When applying these ideas to the usual QCD axion\cite{axionqcd} some subtleties appear. 
The point is that the Lagrangian (\ref{RBmass}) is the dual description  of the usual Higgs effect, 
in which a gauge vector field eats up a pseudoscalar (dual to $B_{\mu\nu}$) and becomes massive. 
On the other hand, the QCD axion gets its mass from the gauge anomaly, and there is no increase 
of the degrees of freedom relative to the massless axion case.   So the correct Higg-type  description 
of this process is when the pseudoscalar is eaten up not by the vector gauge field, but by a composite Chern-Simons three-form
\begin{equation}
\label{qcdform}
C_{\alpha\beta\gamma} \, = \,  {g^2\over 8\pi^2}\, {\rm Tr} \left (A_{[\alpha}A_{\beta}A_{\gamma]}\,  - {3 \over 2}
A_{[\alpha}\partial_{\beta}A_{\gamma]}\right ),
\end{equation}
 where $g$ is the gauge coupling, $A_{\alpha} \, = \, A^a_{\alpha}T^a$ is the gauge field matrix, and $T^a$ are the generators of the group.  Under the gauge transformation, $C$ shifts as 
 \begin{equation}
\label{gaugec}
C_{\alpha\beta\gamma}  \rightarrow  C_{\alpha\beta\gamma} \, + \, \partial_{[\alpha}\Omega_{\beta\gamma]},
\end{equation}
where $\Omega$ is the following  two-form
\begin{equation}
\label{omega}
\Omega_{\alpha\beta}\, = \, A^a_{[\alpha}\partial_{\beta]}\omega^a,
\end{equation} 
and  $\omega^a$ are the $SU(N)$ gauge transformation parameters. 
So the relevant effective dual Langangian for CQD axion is \cite{axi} 
\begin{equation}
\label{bkaction}
 L_{B} \, =  \, \Lambda^4\,  \, \mathcal{K}\left({\mathcal{F} \over \Lambda^2}\right) \, + \,  
 \,   m^2 \, (\partial_{[\alpha}B_{\beta\gamma]}  \, - \, C_{\alpha\beta\gamma})^2 \,,
 \end{equation}
where $\Lambda$ is the QCD scale, $\mathcal{F} \, = \, \epsilon^{\mu\alpha\beta\gamma} \, \partial_{[\mu}C_{\alpha\beta\gamma]}$ 
is the invariant field strength, and 
$\mathcal{K}\left({\mathcal{F} \over \Lambda^2}\right)$ is a nonsingular function of its argument.
After performing the duality transformation to the description in terms of the periodic pseudoscalar axion (call it $\theta_a$), 
the above  function defines the potential for the axion in the following way 
\begin{equation}
\label{connection}
  V(\theta) \, =  \, - \Lambda^2 \int \, {\rm inv}\mathcal{K}'(\theta_a)\, d\theta_a.
\end{equation}
where prime stands for the derivative.  Note that in this description, under  (\ref{gaugec})  $B_{\mu\nu}$ has to shift as  
$B_{\alpha\beta} \, \rightarrow  \, B_{\alpha\beta}  \, + \, \Omega_{\alpha\beta}$, which 
is different from (\ref{shiftB}). 

 Notice that also for the action (\ref{bkaction})  the solution (\ref{bmonopole}) goes through undisturbed, since $dB$ vanishes and we can put $C=0$.  Thus, the massive hair persists in this case. 
 However the detectability of the hair becomes subtle, since the straightforward string coupling  
 (\ref{bstring}) is no longer gauge invariant.  In order to restore the gauge invariance, 
 string should contain a world-sheet scalar degree of freedom (call it $\kappa$)  that would compensate the
 shift. (\ref{bstring}) then gets modified as  
\begin{equation}
\label{bstring1}
q\, \int d^2\sigma \epsilon^{ab} \partial_aX^{\mu}\partial_bX^{\nu} (B_{\mu\nu} \, + \epsilon_{ab}.
\kappa)
\end{equation}
 Non-invariance of the minimal coupling (\ref{bstring}) in the presence of the anomaly, can also be traced to the fact that the axionic string  must inevitably be a boundary of a domain wall. This is trivially seen from (\ref{connection}), which implies that  for any non-trivial $\mathcal{K}$ the symmetry 
 of the QCD vacuum under the continuous shift  of the axion $\theta_a \rightarrow a \, + \,const$ is lifted. 
 The only way to reconcile this lift with the topological obstruction (that $\theta_a$ should wind around the string) is through the formation of the domain wall ending on the string.  Presence of the wall will create an additional interaction between the string and a black hole. 

 The generation of the $\eta'$ meson mass in QCD can also be understood entirely in terms 
 of topological entities\cite{jackiw}, so that (\ref{bkaction}) can be used as an effective theory 
 of the $\eta'$ mass as well. Thus,  all the conclusions about the massive Kalb-Ramond field 
 also apply to $\eta'$-meson, suggesting the possibility for  black holes of 
 having a long range  quantum $\eta'$ hair.

\section{Conclusions}

  In conclusion, black holes can posses  a topologically non-trivial quantum  hair 
  both under the massive spin-2 as well as the massive antisymmetric two-form fields. The latter case is a generalization\cite{massivepseudo}  of the massless axion considered in \cite{pseudo}.
   In both cases the  hair can be detected  at infinity by the generalized Aharovon-Bohm  effect, in which a black hole  passes through a string loop.  The interesting fact is that the effect survives in the limit of an arbitrarily high mass of the corresponding  gauge field.  This gives a prospect of probing a very short distance   physics by large distance observations. In particular, string theory includes plenty of  massive spin-2 particles in form of the Kaluza-Klein excitations, as well as massive $B_{\mu\nu}$ fields. 
 So its in  natural to ask whether the black holes that carry charges under these fields can be detected. 
 
  The fact that black holes may posses a long range quantum hair under the pseudoscalar mesons or 
 the massive spin-2 glueballs and gravitons has  implication for the interaction between the micro black holes and  QCD, fundamental or solitonic  strings. This fact is  of the potential practical interest both for understanding cosmic string - black hole interactions in the early Universe, as well as 
for understanding the collider signatures in theories of  TeV scale quantum gravity, in which 
such object may be accessible at LHC\cite{add}. 
  
 As a byproduct of our analysis, we have shown that  even gauge neutral particles can experience 
 an ordinary Aharonov-Bohm effect in the presence of the photon mass,  provided  certain seemingly-irrelevant  boundary terms are included in the action.  
 
 As usual,  any new type of the charge that can be possessed by the  black hole has a direct relevance 
 for the black hole information loss question.  Massive hair considered above belongs to this
  category, and  corresponding information cannot be lost and must be recovered  after the
 black hole  evaporation.  
   
   {\bf Acknowledgments}

We thank  Gregory Gabadadze, Andrei Gruzinov and Massimo Porrati  for very useful discussions. 
The work  is supported in part  by David and Lucile  Packard Foundation Fellowship for  Science and Engineering, 
and by NSF grant PHY-0245068.

\section{Appendix}

 Let us start from the gauge invariant Pauli-Fierz  Lagrangian supplemented with the boundary term  
 (\ref{total}).
We wish to integrate out $A_{\mu}$ field through its equation of motion (\ref{Aequ}) and write down 
the effective action for the remaining  helicity-2 polarizations of $h_{\mu\nu}$.  We are allowed to do so, since the action is bilinear in $A_{\mu}$. 
By taking a divergence from  the equation (\ref{Aequ})
we get the following constraint on $\hat{h}_{\mu\nu}$ 
 \begin{equation}
\label{hconstraint}
 \partial^{\mu} \partial^{\nu}  \hat{h}_{\mu\nu} \, -  \, \square \hat{h} \, = \, 0,
\end{equation}
which means that $\hat{h}_{\mu\nu}$ is representable in the form 
\begin{equation}
\label{tilderep}
\hat{h}_{\mu\nu} \, = \, \tilde{h}_{\mu\nu} \, + \, \eta_{\mu\nu} {1\over 3} \Pi_{\alpha\beta}\tilde{h}^{\alpha\beta},
\end{equation}
where $\Pi_{\alpha\beta}\, = \, {\partial_{\alpha}\partial_{\beta} \over \square} \, - \, \eta_{\alpha\beta}$ 
is the transverse projector.  $\tilde{h}_{\mu\nu}$ carries two degrees of freedom.
Notice that, since the last term in (\ref{tilderep}) is gauge invariant, under the gauge transformations 
(\ref{gauge}), $\tilde{h}_{\mu\nu}$ shifts in the same way as $\hat{h}_{\mu\nu}$ does. 

Solution of  (\ref{Aequ}) is  
 \begin{equation}
\label{asolution}
A_{\nu} \, = \, - \, {1 \over \square} \partial^{\mu}  (\hat{h}_{\mu\nu} \, - \, \eta_{\mu\nu} \hat{h} )
\, - \, \partial_{\nu} \Theta, 
\end{equation}
where $\Theta$ is arbitrary.  The gauge invariance demands that under the transformations 
(\ref{gauge}), $\Theta$ shifts in the following way
$\Theta \, \rightarrow \, \Theta \, + \, {1 \over \square} \partial_{\alpha}\xi^{\alpha}$.
Substituting (\ref{asolution}) back to (\ref{pfs}) and expressing  
$\hat{h}_{\mu\nu}$ in terms of  $\tilde{h}_{\mu\nu}$ through  (\ref{tilderep}), and choosing 
$\Theta$ appropriately, we get the following effective Lagrangian  


\begin{equation}
\label{effective}
 \tilde{h}^{\mu\nu} \left ( 1 \, - \, {m^2 \over \square} \right )  \mathcal{E}^{\alpha\beta}_{\mu\nu} \tilde{h}_{\alpha\beta}\, + \, 
q\, {1 \over \square} \partial_{[\mu} \partial^{\alpha}\tilde{h}_{\alpha\nu]}   \int d^2\sigma \partial_aX^{\mu}\partial_bX^{\nu} \epsilon^{ab}\delta^4(x - X),
\end{equation}

This action has a solution for which $\tilde{h}_{\mu\nu}$ has a locally pure gauge form 
$\tilde{h}_{\mu\nu} \, = \, \partial_{\mu}\xi_{\nu} \, + \, \partial_{\nu}\xi_{\mu}$, where 
$\xi_{\mu}$ has the form (\ref{ximonopole}),  for which all the locally-observable quantities vanish. 
However, due to the second term it is detectable by the stringy  Aharonov-Bohm effect.


In order to complete the analogy with the massive $U(1)$  Aharonov-Bohm effect, let us follow the same procedure for the Proca action. We start with the action that is the sum of Proca (\ref{procag}) and of the boundary 
coupling (\ref{acoupling}) 
\begin{equation}
\label{procag}
L\, = \, -{1 \over 4} F_{\mu\nu}F^{\mu\nu} \, + \, {1\over 2} \,m^2(\hat{Z}_{\mu} \, + \, \partial_{\mu}a)^2\, +  
\, \beta \int \, dX^{\mu} \partial_{\mu}a,  
\end{equation}
and integrate out the Goldstone boson $a$.  The resulting effective Lagrangian  for $\hat{Z}_{\mu}$ is 
\begin{equation}
\label{procag}
L\, = \,-\,  {1 \over 2} \hat{Z}_{\mu} \left ( \square \, + \, m^2 \right ) \Pi^{\mu\nu} \hat{Z}_{\nu} \, - \, 
\, \beta {1 \over \square} \partial_{\mu} \partial_{\nu} \hat{Z}^{\nu} \int \, dX^{\mu} \delta^4(x-X),  
\end{equation}
In full analogy with massive spin-2 case, the last term produces an Aharonov-Bohm effect for the 
locally-pure-gauge  configuration, with non-trivial winding  $\hat{Z}_{\mu} \, = \, \partial_{\mu}\phi$, 
provided $\beta$ is fractional.

\vspace{0.5cm}   


\end{document}